\documentclass{webofc}
\usepackage[varg]{txfonts}   
\usepackage{hyperref}
\begin{document}
\title{Pseudoscalar pole contribution to the hadronic light-by-light piece of $a_\mu$}
%
%

\author{\firstname{Adolfo} \lastname{Guevara}\inst{1}\fnsep\thanks{\email{
  adguevar@ucm.es}} \and
        \firstname{Pablo} \lastname{Roig}\inst{2}\fnsep\thanks{\email{proig@fis.cinvestav.mx
             }} \and
        \firstname{Juan Jos\'e} \lastname{Sanz Cillero}\inst{1}\fnsep\thanks{\email{jjsanzcillero@ucm.es
             }}
}

\institute{Departamento de F\'isica Te\'orica and UPARCOS, Universidad Complutense de Madrid,
Plaza de las Ciencias 1, 28040 Madrid, Spain
\and
           Centro de Investigaci\'on y de Estudios Avanzados del IPN,
Apartado Postal 14-740, 07000, Ciudad de M\'exico, Mexico
          }

\abstract{%
  We have studied the $P\to\gamma^\star\gamma^\star$ form factor in Resonance Chiral Theory, with 
  $P = \pi^0,\eta,\eta'$, to 
  compute the contribution of the pseudoscalar pole to the hadronic light-by-light piece of the 
  anomalous magnetic moment of the muon. In this work we allow the leading $U(3)$ chiral symmetry
  breaking terms, obtaining the most general expression for the form factor of order $\mathcal{O}(m_P^2)$.
  The parameters of the Effective Field Theory are obtained by means of short distance constraints 
  on the form factor and matching with the expected behavior from QCD. Those parameters that cannot 
  be fixed in this way are fitted to experimental determinations of the form factor within the spacelike 
  momentum region of the virtual photon. Chiral symmetry relations among the transition form factors 
  for $\pi^0,\eta$ and $\eta'$ allow for a simultaneous fit to experimental data for the three mesons. 
  This shows an inconsistency between the BaBar $\pi^0$ data and the rest of the experimental inputs. 
  Thus, we find a total pseudoscalar pole contribution of $a_\mu^{P,HLbL}=(8.47\pm 0.16)\cdot 10^{-10}$
  for our best fit (neglecting the BaBar $\pi^0$ data). Also, a preliminary rough estimate of the 
  impact of NLO in $1/N_C$ corrections and higher vector multiplets (asym) enlarges the uncertainty up to
  $a_\mu^{P,HLbL}=(8.47\pm 0.16_{\rm stat}\pm 0.09_{N_C}{}^{+0.5}_{-0.0_{\rm asym}})$ 
}
\maketitle
\section{Introduction}
\label{intro}

   Nowadays, the anomalous magnetic moment of the muon, $a_\mu$, has been predicted to an outstanding precision of 
   $\mathcal{O}\left[(\alpha/\pi)^5\right]$ for purely electromagnetic effects ($a_\mu^{\rm QED}$)\cite{Kinoshita:2005sm} and to 
   two loops\footnote{The three loop Leading Logarithms are found to give negligible contribution of $\mathcal{O}(10^{-12})$ \cite{PDG}.} 
   precision in electroweak corrections ($a_\mu^{\rm EW}$)\cite{PDG}. This property, which is measured through the decay of 
   muons \cite{Jegerlehner:2009ry}, has been measured to an extraordinary precision of ten significant figures. 
   An interesting point is that the experimental error is orders of magnitude larger than those estimated for $a_\mu^{\rm QED}$ 
   and $a_\mu^{\rm EW}$, however, the uncertainty of the hadronic contributions is of the same order of the 
   experimental error\cite{PDG}. There is an incompatibility among the sum of all contributions with respect to the experiment of $\sim3.5\sigma$, 
   however, there are new experiments that will measure $a_\mu$ with an experimental error reduced at least by a factor 
   4, namely E34 at J-PARC \cite{Iinuma:2016zfu} and muon g-2 at Fermilab \cite{Gohn:2017dsp}.
   If one is to assume the Standard Model (SM) is all that is needed to understand this difference, then there is something 
   about the SM that is not understood; if the assumption is that the difference stems from Beyond Standard Model (BSM) effects, 
   the SM must be further improved to reduce the uncertainties in order to have a more controlled SM background for the search of BSM effects.
   Whatever the case is, the fact that the experimental precision in $a_\mu$ will be further reduced forces theoreticians to 
   give a more precise determination of $a_\mu$.\\
  
 \begin{figure*}[!ht]
 \centering
 \includegraphics[scale=0.32]{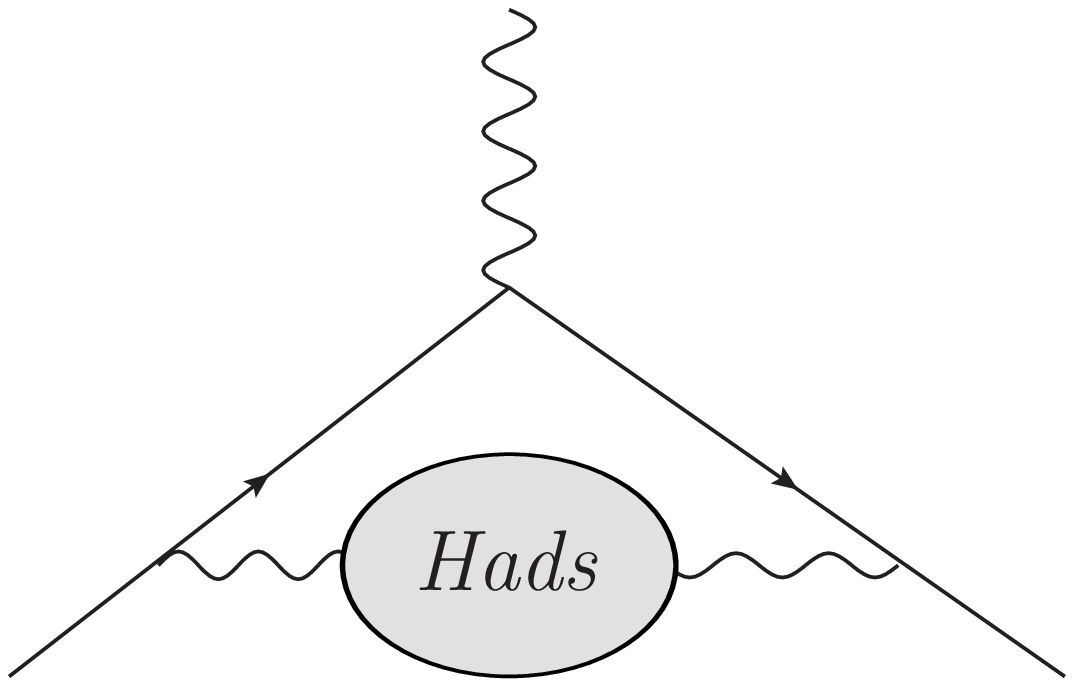}\hspace*{13ex}
 \includegraphics[scale=0.32]{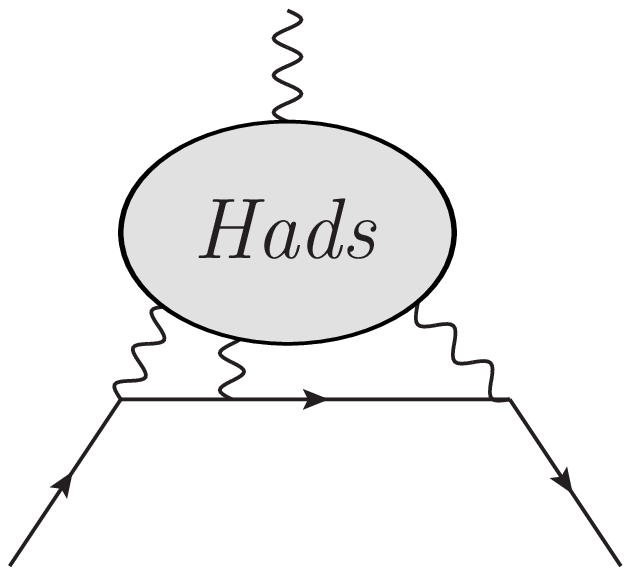}
 \caption{Hadronic contributions to $a_\mu$. The diagram on the left-hand-side represents all contributions from the
 hadronic vacuum to the self energy of the virtual photon, called Hadronic Vacuum Polarization (HVP). The diagram on the 
 right-hand-side represents all contributions from elastic scattering of two photons, called Hadronic Light-by-Light 
 scattering (HLbL).}
 \label{Hadronic_Part}       
 \end{figure*}  
   
   As said above, the main uncertainty comes from hadronic parts. These are contributions involving quarks and their 
   interactions. This can be separated into two pieces, shown in the left-hand-side diagram of Fig. \ref{Hadronic_Part}. 
   The largest of both, $a_\mu^{HVP}$, is the piece stemming from the self energy of virtual photons 
   from the polarization of the hadronic vacuum (HVP) which can, however, be determined through dispersion 
   relations using data available on $R_{\rm had}=\sigma(e^+e^-\to {\rm hadrons})/\sigma(e^+e^-\to\mu^+\mu^-)$.
   The remaining hadronic piece, $a_\mu^{HLbL}$ depends on $\gamma\gamma\to\gamma\gamma$ scattering, which involves strong interactions (HLbL), 
   as shown in the right-hand-side diagram of Fig. \ref{Hadronic_Part}. The latter cannot be obtained in the same 
   manner\footnote{See, however, the outstanding effort done in this direction from \cite{Bern,Mainz}.} as $a_\mu^{HVP}$, 
   and so, has to be obtained either numerically or in a model dependent way. The approach we follow is the latter.\\
   
\begin{figure}[!ht]
    \centering
    \includegraphics[scale=1]{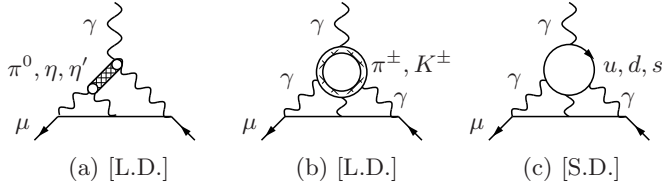}\caption{Leading order contributions in $\alpha$ from HLbL scattering to $a_\mu$. The left-hand-side 
    diagram and its photon momenta permutations give the pseudoscalar pole contribution, $a_\mu^{P,HLbL}$, to $a_\mu$. }\label{HLbL}
   \end{figure}

   The HLbL can be divided in three parts, which are the leading order parts of the HLbL piece \cite{Jegerlehner:2009ry} shown in Fig. \ref{HLbL}. 
   The sum of diagrams \ref{HLbL} (b) and \ref{HLbL} (c) are one order of magnitude suppressed as compared to the pseudoscalar exchange 
   shown in diagram \ref{HLbL} (a). We focus on the pseudoscalar pole contribution to the HLbL piece, $a_\mu^{P,HLbL}$, which 
   in order to be fully described needs only the Transition Form Factor (TFF), $\mathcal{F}_{P\gamma^\star\gamma^\star}(q^2,p^2)$. 
   As it has been shown in ref \cite{Knecht:2001qf}, this TFF gives most of its contribution to $a_\mu$ at Euclidian squared photon momenta $\lesssim 1$ GeV$^2$. 
  Therefore, it will be saturated mainly with the lightest resonant part of the (TFF) and higher energies effects will be suppressed. 
   To compute the TFF we rely on $\chi$PT \cite{ChPT,WZW} extended to include the lightest multiplet of meson resonances \cite{RCT} (Resonance 
   Chiral Theory, R$\chi$T). 
  Instead of using the complete basis of VVP and VJP operators for resonances \cite{Kampf:2011ty}, we will rely on the simpler 
  basis given in \cite{RuizFemenia:2003hm} since, for describing vertices involving only one pseudo-Goldstone both are 
  equivalent \cite{Roig:2013baa}. We will use the Lagrangian in ref. \cite{Kampf:2011ty} to include effects due to pseudoscalar resonances. 
  The novelty in our approach is that we account for all the leading order terms that 
  break explicitly chiral symmetry, which enter as corrections in powers of the squared pseudo-Goldstone bosons, $m_P^2$. 
  Here we will only present those operators necessary for a consistent description of the NLO terms in $m_P^2$.
   
 \section{Flavor {$U(3)$} breaking}
 \label{sec-1}
  For the complete description of the full base of operators used \cite{RCT,WZW,Bijnens:2001bb,RuizFemenia:2003hm,Cirigliano:2006hb,Kampf:2011ty}, we refer to \cite{Guevara:2018rhj}.
  To consistently include all terms breaking $U(3)$, an $\mathcal{O}(p^6)$ odd-intrinsic Lagrangian without resonances must be considered for R$\chi$T \cite{Bijnens:2001bb}. 
  In addition to the Wess-Zumino-Witten term \cite{WZW}, the relevant operators here are 
  \begin{eqnarray}\label{p6ChPT}
   \hspace*{20ex}\mathcal{O}_7^W&=&i \epsilon_{\mu\nu\alpha\beta} \langle \chi_- f_+^{\mu\nu} f_+^{\alpha\beta}\rangle,\nonumber\\
   \mathcal{O}_8^W&=&i \epsilon_{\mu\nu\alpha\beta} \langle \chi_-\rangle \, \langle  f_+^{\mu\nu} f_+^{\alpha\beta}\rangle,\nonumber\\
   \mathcal{O}_{22}^W&=&i \epsilon_{\mu\nu\alpha\beta} \langle u^\mu \{ \nabla_\rho f_+^{\rho\nu}, f_+^{\alpha\beta}\}\rangle.
  \end{eqnarray}
  A correction to the vector resonance-photon coupling will be given by the interaction\footnote{This interaction term is 
  the only single-trace operator $\mathcal{O}(m_P^2)$ from those given in \cite{Cirigliano:2006hb}.}
  \begin{equation}
   \mathcal{L}_{VJ}=\frac{\lambda_V}{\sqrt{2}}\langle V_{\mu\nu}\{f^{\mu\nu}_+,\chi_+\}\rangle.
  \end{equation}
  There is also a correction to the mass of the vector resonances from V-V interactions\footnote{This term generates a mass splitting effect in the 
  nonet of resonances, inducing an explicit $U(3)$ breaking effect.}
  \begin{equation}
   \mathcal{L}_{VV}=-e_m^V\{V_{\mu\nu}V^{\mu\nu}\chi_+\}.
  \end{equation}
  As a result, the masses of the vector resonances are given by
  \begin{equation}\label{VMasses}
   M_\rho^2=M_\omega^2 = M_V^2 - 4 e_m^V m_\pi^2,\hspace*{10ex} M_\phi^2=M_V^2 - 4 e_m^V \Delta_{2K\pi}^2,
  \end{equation}
  where $\Delta_{2K\pi}^2=2m_K^2-m_\pi^2$ and $M_V$ is the mass associated with the vector nonet in the chiral and large $N_C$ limits. 
  The TFF is defined in ref. \cite{Guevara:2018rhj} along with 
  the short distance constraints relating parameters of the model. Its complete expression before 
  imposing such constraints can be found therein.
  \section{Transition Form Factor}\label{sec:TFF}
   By imposing the short-distance behavior 
   \begin{equation}\label{SDc}
    \lim_{q^2\to\infty}\mathcal{F}_{P\gamma^\star\gamma^\star}(q^2,q^2)=\mathcal{O}(q^{-2}){\rm\quad\quad and\quad\quad}
    \lim_{q^2\to\infty}\mathcal{F}_{P\gamma\gamma^\star}(0,q^2)=\mathcal{O}(q^{-2}),
   \end{equation}
    we find some relations among the parameters of the theory that simplify the expression of the TFF. For $\pi^0$ we find\footnote{
    $d_2^\star$ and $d_{123}^\star$ have been defined to account for pseudoscalar resonance effects.}
   \begin{equation}\label{SimppiFF}
    \mathcal{F}_{\pi\gamma^\star\gamma^\star}(q_1^2,q_2^2)=\frac{32\pi^2m_\pi^2F_V^2d_{123}^\star -  N_C M_V^2M_\rho^2
    }{12 \pi ^2 F_\pi D_\rho(q_1^2)
    D_\rho(q_2^2)},
   \end{equation}
   where $D_R(s)=M_R^2-s$ with the resonance masses given by (\ref{VMasses}), $d_{123}^\star$ is a free parameter and $F_\pi$ 
   is the $\pi$ decay constant. Analogously, the simplified expression for the TFF of the $\eta$ after imposing eq. (\ref{SDc}) is given by
    \begin{eqnarray}\label{SimpetaFF}
    \mathcal{F}_{\eta\gamma^\star\gamma^\star}(q_1^2,q_2^2)&=&\frac{1}{12\pi^2F D_\rho(q_1^2) D_\rho(q_2^2)D_\phi(q_1^2) D_\phi(q_2^2)}\times
    \\
     &&
     \hspace*{-2cm}\left\{-\frac{N_C M_V^2}{3     }\left[5C_q M_\rho^2D_\phi(q_1^2) D_\phi(q_2^2) - \sqrt{2}C_s M_\phi^2 D_\rho(q_1^2) D_\rho(q_2^2)\right]\right.
     \nonumber\\
     &&
     \hspace*{-2cm}+\frac{32\pi^2F_V^2d_{123}^\star m_\eta^2}{3}\left[(5C_q D_\phi(q_1^2) D_\phi(q_2^2)-\sqrt{2}C_s D_\rho(q_1^2) D_\rho(q_2^2)\right]
   \nonumber\\
   &&
   \hspace*{-2cm}\left.-\frac{256\pi^2F_V^2d_{2}^\star}{3}\left[(5C_q\Delta_{\eta\pi}^2 D_\phi(q_1^2) D_\phi(q_2^2)+\sqrt{2}C_s\Delta_{2K\pi\eta}^2 D_\rho(q_1^2) D_\rho(q_2^2)\right]\right\},\nonumber
 \end{eqnarray}
 where $d_2^\star$ is another free parameter and $C_{q/s}$ are the $\eta-\eta'$ mixing parameters. One can get the $\eta'$-TFF from eq. (\ref{SimpetaFF}) by making 
 $C_q\to C_q'$, $C_s\to -C_s'$ and $m_\eta\to m_{\eta'}$.\\
 
 As it can be seen from the previous expressions, we were not able to match the precise QCD behavior \cite{BL,Nesterenko}
 $\lim_{q^2\to\infty}\mathcal{F}_{P\gamma^\star\gamma^\star}(q^2,q^2)\propto1/q^2$, which will be taken into account as a systematic uncertainty. 
 This implies an underestimation of the form factor, and thus will enter as an asymmetric error.
 However, it is worth to mention that the correct behavior in $q^2$ with approximately the correct coefficient is achieved for the singly virtual TFF, 
 $\lim_{q^2\to\infty}\mathcal{F}_{P\gamma^\star\gamma}(q^2,0)\approx2F/q^2$, after the fit to experimental data.

\section{Pole contribution to $a_\mu^{HLbL}$}
\label{sec:amu}
The parameters $d_2^\star$, $d_{123}^\star$, $M_V$, $e_m^V$ and the $\eta-\eta'$ mixing parameters are all fitted to experimental data from  
BaBar (\cite{BaBarpi} for $\pi$-TFF and \cite{BABAR:2011ad} for $\eta^{(\prime)}$-TFF), CLEO \cite{CLEO} and CELLO \cite{CELLO}, 
LEP \cite{Lep} and Belle \cite{Belle}. The complete details of the fits can be seen on ref. \cite{Guevara:2018rhj}. To compute $a_\mu^{P,HLbL}$
we rely on the loop integral representation of ref. \cite{Knecht:2001qf}. The $a_\mu^{P,HLbL}$ contribution is obtained using a MonteCarlo run 
based on a normal distribution for the fit parameters, including correlations\footnote{By making the same calculation for the TFF in the chiral 
limit, we obtain $a_\mu^{P,HLbL}=8.27\cdot10^{-10}$, where the change is essentially 
given by the $\eta$ contribution. We note that, however, we keep the physical masses for the integration kernels.},
\begin{equation}\label{amuP}
 a_\mu^{P,HLbL}= (8.47 \pm 0.16)\cdot10^{-10}.
\end{equation}
Also, the separate parts of the three pseudo-Goldstone mesons are 
computed in the same way, giving
\begin{subequations}
 \begin{align}
  a_\mu^{\pi^0,HLbL}&=(5.81\pm0.09)\cdot10^{-10},\\
  a_\mu^{\eta,HLbL}& =(1.51\pm0.06)\cdot10^{-10},\\
  a_\mu^{\eta',HLbL}&=(1.15\pm0.07)\cdot10^{-10}.
 \end{align}
\end{subequations}
 Notice that our result for the $\pi^0$ contribution is in agreement within $1.8\sigma$ with that obtained using dispersion relations 
 \cite{Hoferichter:2018kwz}.\\

 Note that by taking the uncorrelated error as a sum in quadratures from the three previous values one gets a smaller uncertainty 
 ($\sim \pm0.13\cdot10^{-10}$) than that given in eq. (\ref{amuP}). This is due to correlations among the three results, since the 
 total result is obtained performing the simultaneous integral of the three contributions. In Table \ref{tab-1} we compare our 
 result with some previous determinations of $a_\mu^{P,HLbL}$.

 \begin{table}[!ht]
 \centering
 \caption{Comparison of different predictions for the pseudoscalar-pole contributions to $a_\mu^{HLbL}$. The `sta' error 
is due to fit uncertainties. The details on the other errors are given in section \ref{error}}
 \label{tab-1}       
 \begin{tabular}{lll}
 \hline
 Reference & $10^{10}\, \cdot \,  a_\mu^{P,HLbL}$ \\\hline
     Knecht and Nyffeler \ (2002)  \cite{Knecht:2001qf} &\hspace*{0.2cm} 8.3 \ \ $\pm$ \ 1.2\\
     Hayakawa and Kinoshita \ (2002)  \cite{Hayakawa:1997rq} &\hspace*{0.2cm}  8.3 \ \ $\pm$ \ 0.6 \\
     Bijnens, Pallante and Prades \ (2002) \cite{Bijnens:2001cq} &\hspace*{0.2cm}  8.5 \ \ $\pm$ \ 1.3\\
     Erler and Toledo S\'anchez\ (2006) \cite{Erler:2006vu} &\hspace*{0.2cm}  13.7  \ \ $^{+2.7}_{-1.5}$
     \\
     Roig, Guevara and L\'opez Castro \ (2014) \cite{Roig:2014uja} &\hspace*{0.2cm}  8.60 \ $\pm$ \ 0.25\\
     Masjuan and S\'anchez-Puertas \ (2017) \cite{Masjuan:2017tvw} &\hspace*{0.2cm}  9.4 \ \ $\pm$ \ 0.5\\
     Czy\.z, Kisza and Tracz \ (2018) \cite{Czyz:2012nq} &\hspace*{0.2cm}  8.28\  $\pm$ \ 0.34\\
     This work \cite{Guevara:2018rhj}
     &\hspace*{0.2cm}  $8.47   \pm  0.16_{\rm sta} \pm 0.09_{N_C}{}^{+0.5}_{-0.0_{\rm asym}})$\\
     \hline
     
 \end{tabular}
 \end{table}

 \section{Further error analysis}\label{error}
 
 Our previous result does not take into account NLO effects in the $1/N_C$ expansion in which R$\chi$T relies. However, 
 these effects can be estimated by taking into account how the main contribution at this order to the $\rho$ meson propagator, 
 namely the $\pi\pi$ and $K\overline{K}$ loops, affect the result for $a_\mu^{P,HLbL}$. This loop effects are included by 
 changing the denominator of the propagator \cite{GomezDumm:2000fz}

  \begin{equation}
  M_\rho^2 - q^2 \,\, \longrightarrow\,\, 
  M_\rho^2-q^2 +\frac{q^2M_\rho^2}{96\pi^2F_\pi^2}\left(A_\pi(q^2)+\frac{1}{2} A_K(q^2) \right),
  \label{eq:prop-NLO}
 \end{equation}
 where
 \begin{equation}
  A_P(q^2)\, =\,  \ln \frac{m_P^2}{M_\rho^2}+8\frac{m_P^2}{q^2}-\frac{5}{3}+\sigma_P^3(q^2)\ln\left(\frac{\sigma_P(q^2)+1}{\sigma_P(q^2)-1}\right),
 \label{eq:selfenergy-loop}
 \end{equation}
 being $\sigma_P(q^2)=\sqrt{1-\frac{4m_P^2}{q^2}}$. Note that $A_P(q^2)$ is real for $q^2< 4 m_P^2$. Thus, the $\rho$~propagator 
 provided by eq. (\ref{eq:prop-NLO}) is real in the whole space-like region $q^2<0$. This gives a total contribution to the uncertainty 
 \begin{equation}
  {\Delta a_\mu^{P,HLbL}}_{1/N_C} = \pm0.09\cdot10^{-10}
 \end{equation}
 
 As said in section \ref{sec:TFF}, the form factor is underestimated due to the incorrect asymptotic behavior of the doubly off-shell form factor. 
 This was previously noted by \cite{Knecht:2001qf}, where it was pointed out that the correct asymptotic behavior can be recovered considering 
 two vector meson resonance multiplets instead of just one. Therefore, in order to give an estimation of the uncertainty stemming from this incorrect limit 
 we compute $a_\mu^{P,HLbL}$ considering two multiplets and compare the result with that using one multiplet, both in the chiral limit. The values for 
 the fit parameters are kept identical for both estimations of $a_\mu^{P,HLbL}$. For a more detailed description of the form factor and the short distance 
 constraints obtained among parameters we refer to \cite{Guevara:2018rhj}. This uncertainty is estimated to be
 \begin{equation}
  {\Delta a_\mu^{P,HLbL}}_{\rm asym} = ^{+0.5}_{-0.0}\cdot10^{-10}.
 \end{equation}
\section{Conclusions}

We have given a more accurate description of the TFF within the framework for R$\chi$T, including terms up to order $m_P^2$ for the first time 
in a chiral invariant Lagrangian approach. This led to computing a more theoretically precise contribution from the P-pole to $a_\mu$, 
with an improved precision. By looking at the difference of our result with that using the TFF in the chiral limit ($0.20\cdot10^{-10}$)  
it is evident that further orders in $m_P^2$ will be negligible. Future works will be directed towards reducing the uncertainty from the asymptotic 
limit, which is the largest one. A way to reduce such uncertainty could be by taking into account data form doubly off-shell TFF such as that 
given by BaBar for the $\eta'$-TFF \cite{BaBar:2018zpn}. Considering all possible contributions to the error we get
\begin{equation}
 a_\mu^{P,HLbL} \,\, =\,\,  (\, 8.47\, \pm\, 0.16_{\rm sta}\, \pm \, 0.09_{ 1/N_C}\, {}^{+0.5}_{-0}{}_{\rm asym}\, )\, \cdot\, 10^{-10},\, 
\end{equation}
 where the first error (sta) comes from the fit of the TFF, the second from possible $1/N_C$ corrections and the last from the wrong asymptotic 
 behavior estimated through the effects of heavier resonances in the TFF.
 \section{Acknowledgements}
 This work was supported by CONACYT Projects No. FOINS-296-2016 ('Fronteras de la Ciencia')
and 250628 ('Ciencia Básica'), by the Spanish MINECO Project FPA2016-75654-C2-1-P
and by the Spanish Consolider-Ingenio 2010 Programme CPAN (CSD2007-00042). A. G.
acknowledges CONACYT for the support 'Estancia Posdoctoral en el Extranjero'. 
J.J.S.C. would like to thank Z.H. Guo for discussions
on the $\eta-\eta'$ mixing. P.R. acknowledges discussions on the short distance constraints with
Bastian Kubis, Andreas Nyffeler, Hans Bijnens and Gilberto Colangelo during the 'Muon
g-2 Theory Initiative Hadronic Light-by-Light working group workshop' held at Univ. of
Connecticut, 12-14 March 2018.
%

%

%
%

\end{document}